\documentclass[twocolumn]{article}
\usepackage{amsmath}
\usepackage{amssymb}
\usepackage{graphicx}
\usepackage{epsfig}
\usepackage{color}

\begin{document}

\title{Anisotropic generalization of Stinchcombe's solution
 for conductivity of random resistor network on a Bethe lattice}
\author{F. Semeriyanov, M. Saphiannikova, and G. Heinrich \\
Leibniz Institute of Polymer Research Dresden, Hohe
str. 6, 01069 Dresden, Germany}
\date{\today}
\maketitle
\begin{abstract}
Our study is based on the work of Stinchcombe [1974 \emph{J. Phys.
C} \textbf{7} 179] and is devoted to the calculations of average
conductivity of random resistor networks placed on an anisotropic
Bethe lattice. The structure of the Bethe lattice is assumed to
represent the normal directions of the regular lattice. We calculate
the anisotropic conductivity as an expansion in powers of inverse
coordination number of the Bethe lattice. The expansion terms
retained deliver an accurate approximation of the conductivity at
resistor concentrations above the percolation threshold. We make a
comparison of our analytical results with those of Bernasconi [1974
\emph{Phys. Rev. B} \textbf{9} 4575] for the regular lattice.
\end{abstract}

\section{Introduction}
The random percolation theory due to Broadbent and Hammersley
\cite{BroadbentHammersley} is too simple to explain the great
variety of percolation phenomena. One confronts complexity of real
systems with both correlations and anisotropy playing important
role. The motivation for our study is to understand better the
nature of the anisotropy in electrical conductivity of percolating
systems. This is approached by means of the random resistor network
(RRN) originally proposed by Kirkpatrik \cite{Kirkpatrick}. The
resistor networks can be associated with the networks of saddle
points in the conductivity profile of high-contrast systems as
proved in \cite{Borcea}. Besides conductivity, RRN has been used to
predict magnetic properties of materials \cite{XuHui} and even to
estimate sample destruction under critical mechanical stress
\cite{AcharyyaChakrabarti1}. In the past there have been several
propositions of the anisotropic percolation theories based on an
assumption that the lattice bond occupation probability is
\emph{dependent} on the spatial orientations
\cite{Turban,Blane,Friedman,RednerStanley,Gavrilenko}. As a result,
these theories associate the direction with the percolation
threshold too, which, however, may not be true in the case of
composites filled by long sticks. As found in the Monte Carlo
simulations \cite{Balberg}, the percolation threshold measured in
the directions parallel and normal to the direction of the average
orientation, merge to a single value in the limit of infinitely
large length of the sticks. Another class of anisotropic percolation
theories \cite{Bernasconi, Straley} assumes the occupation
probability to be \emph{independent} of the spatial directions,
whereas the local conductivity is assumed to be a
direction-dependent property. Unfortunately, all these theories
cannot describe a peculiar phenomenon observed in geophysics: The
Earth mantle exhibits the scale-dependent behavior of its
conductivity anisotropy, viz. its macroscopic anisotropy is much
more pronounced than the microscopic one. This seems to be an
indication of the fractal nature of the geological networks, see
\cite{Bahr1, Bahr} and references therein. The latter together with
the fact that Earth drainage networks have a tree-like topology
\cite{TurcotteNewman}, makes us to believe that the topology of the
resistor network behind the conductive property of the Earth mantle
may also be tree-like in nature.

Using the exact Bethe lattice solution obtained by Stinchcombe
\cite{Stinchcombe} we propose an anisotropic RNN model that combines
both the advantage of the recursive structure of a tree and the
notion of a direction. In the present contribution, it will be
demonstrated that the latter, being geometrically clear on a regular
lattice, can be associated with Bethe lattice as well.
Unfortunately, the original paper \cite{Stinchcombe} has given rise
to a highly puzzling and controversial issue \cite{Straley,Straley1}
regarding the critical exponent 2 being close to the real value in
3D instead of the expected mean-field value of 3 \cite{deGennes}. To
make the situation even more confusing, it was observed
\cite{Heibaetal,Sahimi} that Stinchcombe's solution serves as a very
good approximation to the macroscopic conductivity of the resistor
network on the regular 3D lattice. As highlighted by the present
state of understanding of this problem \cite{Sahimi1}, those two
facts are just the matter of mere coincidence.

To refute this strongly negative disposition, we want to show that
the correlations captured by the Bethe lattice, being controlled by
the coordination number $z$, are sufficient to produce a very good
fit to the exact solution of Bernasconi \cite{Bernasconi} for the
anisotropic RNN on the regular lattice. The latter applies when the
occupation probability is \emph{well above} the critical point. At
the same time, it is well known that the correlations captured are
not sufficient to obtain the right critical exponents.

\begin{figure}
\epsfxsize=3.4in \epsffile{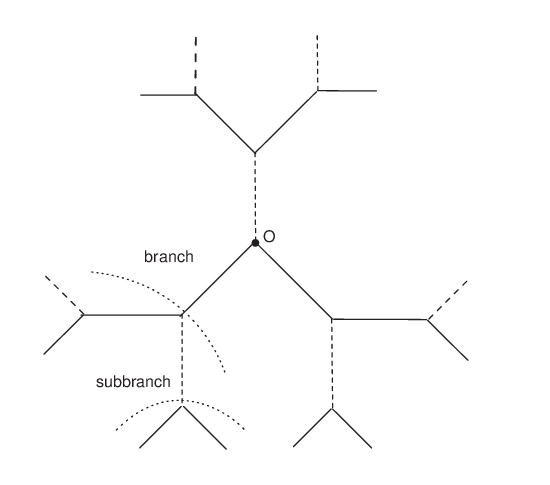}%
\caption{Anisotropic Bethe lattice of coordination number $z=3$ with
two kinds of bonds, $\alpha$ and $\beta$, depicted by solid and
dashed lines. The center O, referred to as origin, is where
$n_{\alpha}$ and $n_{\beta}$ branches are connected by their root
bonds of $\alpha$ and $\beta$ kind, respectively. $z = n_{\alpha} +
n_{\beta}$. Each branch is made of $z-1$ subbranches connected
together by their root bonds.} \label{FigBetheLettice}
\end{figure}

Technically, we generalize the Stinchcombe's calculation to the case
of the anisotropic Bethe lattice, see Fig \ref{FigBetheLettice}.
Besides absence of closed loops, this structure has a special
feature of being anisotropic at each node. Specifically, there are
$n_{\alpha}$ bonds of $\alpha$ kind and $n_{\beta}$ bonds of $\beta$
kind connected at each branching point. At the same time, their
total sum at a node is equal to a constant number $z$ referred to as
the coordination number of the lattice. We would like to stress that
there is a large difference between the finite Bethe lattice, known
also as the Cayley tree, and the infinite lattice with the surface
sites neglected by definition, the difference being carefully
discussed by Gujrati and Bowman \cite{GujratiBowman}.

The outline of the paper is as follows: in section II we present the
model and mathematical formulation of the problem, in section III we
present the main results, in section IV we describe our verification
of the theory with the exact solution of Bernasconi, in section V we
describe the connection with experiment, in section VI we provide a
discussion and conclusions. Appendices A-C provide the details of
our computation.

\section{Model}
Unlike several previous anisotropic percolation theories
\cite{Turban,Blane,Friedman,RednerStanley,Gavrilenko} based on the
model of different probabilities of filling for two types of lattice
bonds, we consider the distribution of resistors to be isotropic.
However we make the local conductivities of the network elements on
this special Bethe lattice to be a 'direction'-dependent, i.e. equal
to $\sigma_{\alpha}$ and $\sigma_{\beta}$ for $\alpha$ and $\beta$
occupied bonds, respectively. The lattice itself is considered to be
non-conductive. Thus, the resistors, associated with occupied bonds
of the lattice, are the only conductive objects forming a tree-like
network. In the mathematical form, the local conductivity
distribution function is written as follows:
\begin{equation}
    g_{\alpha}(\sigma) = p\,\delta(\sigma-\sigma_{\alpha}) +
    (1-p)\delta(\sigma),
\end{equation}
where $p$ is the bond occupation probability, common for the bonds
of both types. Given $p$ and the conductivities of network elements,
$\sigma_{\alpha}$ and $\sigma_{\beta}$, we compute the average
conductivity of the network connected to a constant potential source
at the origin and grounded at infinity. The question how to perform
configurational averages turns out to be a difficult one.

A starting point of the present development is the observation that
the percolation threshold is given by the usual equation
\cite{Stauffer}:
\begin{equation}
    p_{c}=1/(z-1), \label{pc}
\end{equation}
as its derivation does not require considerations of conductivity as
such. This is the consequence of the occupation probability common
for the bonds of both kinds.

Further, we define the probability distribution functions, $\phi
_{\alpha}(b)$ and $\phi _{\beta}(b)$, for the average conductivity
of a branch being some value $b$,
\begin{equation}\label{phi_norm_cond}
    \int_{0}^{\infty} \phi_{\alpha}(b) db = 1.
\end{equation}
Those functions measure the contribution from averaging over
ensemble sampled by resistor permutations, so that the average
branch conductivity is given by
\begin{equation}
    \overline{b_{\alpha}} = \int_{0}^{\infty} b
    \phi_{\alpha}(b) db \label{b_av_def_1}
\end{equation}

Here, the symmetry $\alpha \leftrightarrow \beta$ holds for all
quantities. Note that we specified in (\ref{phi_norm_cond}) and
(\ref{b_av_def_1}) the $\alpha$-components, only, for the sake of
brevity. The second equation is obtained readily using the $\alpha
\leftrightarrow \beta$ interchange. This convention is followed
everywhere in the text.

The average conductivity of a part of the tree consisting of
$n_{\alpha}$ branches connected at the origin in parallel is
\begin{equation}
    \overline{\sigma_{\alpha}} = n_{\alpha}\, \overline{b_{\alpha}}.
    \label{sigma_avg}
\end{equation}
As an example, one can take $n_{\alpha}=n$, and $n_{\beta}=z-n$,
which leads to $n-1$ and $z-n$ of $\alpha$- and $\beta$-subbranches,
respectively, for the $\alpha$ branch (see Fig.
\ref{FigBetheLettice}). In order to compute $\phi_{\alpha}(b)$ and
$\phi_{\beta}(b)$ we use the algorithm of Ref. \cite{Stinchcombe}
modified to account for the lattice anisotropic structure, detailed
calculation is given in Appendices A-C.

\section{Results}
The analytical solutions have been obtained for the two cases: (I)
for the case of infinitely large coordination number, $z \rightarrow
\infty$ and (II) near the percolation threshold, $p \approx p_{c}$.
In both cases the solution is represented in the form of a Taylor
expansion in terms of the small parameters, $p_{c}=(z - 1)^{-1}$ and
$\epsilon = (p-p_{c})/p_{c}$, respectively.

In the first case we obtain (see Appendix B for details)
\begin{equation}
    \overline{b_{\alpha}}_{\text{(I)}} = -\sigma_{\alpha}
    \left( -\frac{p \sigma_{\beta} \Delta }
    {\sigma_{\alpha}p_{c}+\sigma_{\beta} \Delta} +
    \frac{\Delta}{p}
    \sum_{k=2}^{\infty}G^{(k)} \right), \label{sigma_final}
\end{equation}
where
\begin{eqnarray}
  G^{(2)} &=& \frac{p_{c}^{2}}{p^2}\,\Delta \,s, \nonumber \\
  G^{(3)} &=& \frac{p_{c}^3}{p^5}\,\Delta^2 s [ p(2p-1)+3\Delta s
  ], \nonumber \\
  G^{(4)} &=& \frac{p_{c}^{4}}{p^6}\,\Delta^{2}s \left[
                     3s^{2}\Delta - 2sp
                     \textcolor[rgb]{1.00,1.00,1.00}
                     {\frac{\frac{}{}}{\frac{}{}}}
                     \right. \nonumber \\
  &&+\, \Delta (1-3p+3p^{2}) \nonumber \\
  &&\left. +\, 10\Delta^{2}s\left( \frac{2p-1}{p} \right)
  + 15\Delta^{3}s^2 \frac{1}{p^2} \right], \nonumber \\
  G^{(k)} &=& \text{O}(p_{c}^{k}), \nonumber  \\
  \Delta &=& p-p_{c}, \>\>\>\> \text{and} \>\>\>\> s=1-p.
  \label{Stinch}
\end{eqnarray}
This equation gives the average conductivity of a branch starting
from the $\alpha$-bond connected to a potential difference between
the node at its root and the nodes at infinity. The first term is
the conductivity of the infinitely branched Bethe lattice, while the
summation over $G$-s represents the corrections up to and including
$(z-1)^{-4}$ order. To understand the differences with the isotropic
case we reproduce the expression obtained by Stinchcombe
\cite{Stinchcombe}:
\begin{equation}
    \overline{b}_{\text{iso(I)}} = \sigma \left( \Delta -
    \sum_{k=2}^{\infty}G^{(k)} \right). \label{Stinch_form}
\end{equation}
with $G^{(2)}$ and $G^{(3)}$ the same as in (\ref{Stinch}), but
$G^{(4)}$ being given by
\begin{eqnarray}
   G^{(4)} &=& \frac{p_{c}^{4}}{p^6}\,\Delta^{2}s\left[
                    3s^{2}\Delta - 2sp
                    \textcolor[rgb]{1.00,1.00,1.00}
                    {\frac{\frac{}{}}{\frac{}{}}}
                    \right. \nonumber \\
  &&+\, \Delta (1-3p+3p^{2}) \nonumber \\
  &&\left.+\, 5\Delta^{2}s\left( \frac{2p-1}{p} \right)
  + 15\Delta^{3}s^2 \frac{1}{p^2} \right]. \label{Stinch1}
\end{eqnarray}
In the isotropic case, $\sigma_{\alpha} = \sigma_{\beta} = \sigma$,
one finds that Eq. (\ref{Stinch_form}) is different from our result
by the factor $\Delta/p$ before the summation. In addition, Eq.
(\ref{Stinch1}) contains the factor 5 in front of $\Delta^{2}s\left(
\frac{2p-1}{p} \right)$ different to 10 we have. We want point out
that these discrepancies play a minor numerical role in the
isotropic case as will be demonstrated in the next Section.

We now calculate the critical exponents and the anisotropy near the
percolation threshold, $p \approx p_{c}$. The details of the
calculation are shown in Appendix C. Close to the critical point the
integer numbers $n_{\beta}$ and $n_{\alpha}$ are set by
\begin{equation}
    n_{\alpha} \sigma_{\beta} = n_{\beta} \sigma_{\alpha},
    \label{n_sigma_results}
\end{equation}
received from the symmetry considerations, eq. (\ref{n_sigma}).
Qualitatively, this can be explained as follows: Bethe lattice, with
its origin O representing a point inside the sample, has the
branching topology of the infinite cluster. Suppose the system is
just above $p_{c}$. Although the formation of the spanning cluster
is a topological concept, it is qualitative clear that that the
physics at $p_{c}$ is dominated by singly connected bonds that are
present on all length scales, which made Skal and Shklovskii
\cite{SkalShklovski} and de Gennes \cite{deGennes1} to postulate
that within each box of size of the correlation length $\xi$ there
is only one chain of bonds that connects its opposite edges, see
also \cite{Stauffer}. Thus, it is possible to associate the average
direction of these chains with the average direction of the infinite
cluster, which should be the direction where the resistance to
current is minimal. For the case when the occupation probability $p$
is the same in all directions, the direction dependent percolation
probability can only be achieved if the fraction of bonds of one
kind is larger than another. Indeed, the isotropic percolation
probability $P = 1 - R^{z}$, where $R<1$ is the probability to have
the finite cluster \cite{Stauffer}, can be generalized to the
anisotropic one, $P_{\alpha}=1-R^{n_{\alpha}}$, which gives
$P_{\alpha}>P_{\beta}$ if $n_{\alpha}>n_{\beta}$. This leads to the
following conditions:
\begin{eqnarray}
    n_{\alpha} =
    \frac{z\sigma_{\alpha}}{\sigma_{\alpha}+\sigma_{\beta}}, \>\>\>
    n_{\beta} = z - n_{\alpha}. \label{n_alpha_n_beta_new}
\end{eqnarray}
Thus, $n_{\alpha}$ and $n_{\beta}$ are fixed by the local
conductivities.

Returning for a moment to the previous case, we note that the Bethe
lattice topology should be intact on the change of $p$. Thus, the
condition (\ref{n_alpha_n_beta_new}) has also to be applied above
the critical point to obtain
$\overline{\sigma_{\alpha}}_{\text{(I)}}$ from eqs.
(\ref{sigma_avg}) and (\ref{sigma_final}):
\begin{equation}
    \overline{\sigma_{\alpha}}_{\text{(I)}} =
    \frac{z\sigma_{\alpha}}{\sigma_{\alpha}+\sigma_{\beta}}
    \overline{b_{\alpha}}_{\text{(I)}} \label{sigma_final_new}
\end{equation}

In the critical region, we investigate the anisotropy ratio of the
network conductivities and relate this to the experimental quantity
$\overline{\sigma_{||}} / \overline{\sigma_{\bot}}$, where
$\overline{\sigma} _{||,\bot}$ are the bulk conductivities parallel
and normal to the direction of an applied voltage. According to Skal
and Shklovskii \cite{SkalShklovski},
\begin{equation}
    \overline{\sigma_{||}} / \overline{\sigma_{\bot}}  \simeq 1 +
    (p-p_{c})^{\lambda(d)}, \label{skal_shklovskii}
\end{equation}
where $\lambda(d)$ is a critical exponent determined by $d$ - the
dimensionality of a problem.

Straley \cite{Straley}, who first studied the conductivity exponent
on the anisotropic Bethe lattice near the percolation threshold,
obtained the anisotropy critical exponent $\lambda=1$. Sarychev and
Vinogradov\cite{SarychevVinogradov} using the renormalization group
theory and computer simulations found that $\lambda(2) = 0.9 \pm
0.1$ and $\lambda(3) = 0.3 \pm 0.1$ for 2D and 3D, respectively.
Carmona and Amarti\cite{CarmonaAmarti} deduced from experimental
data for short carbon fiber reinforced polymers that $\lambda(3)
\approx 0.4$. The details of our computation are given in Appendix
C. Our final result (\ref{sigma_avg_exp_crit1}), written in a more
concise form, is given by
\begin{equation}
    \overline{\sigma_{\alpha}}_{\text{(II)}} =
    0.762 \frac{z}{z-2} \frac{2\sigma_{\alpha}\sigma_{\beta}}
    {\sigma_{\alpha}+\sigma_{\beta}}
    \epsilon^2 + O(\epsilon^{3}), \label{b_near_pc}
\end{equation}
where $\epsilon = (p - p_{c}) / p_{c}$. We also calculated the
average anisotropy ratio near the percolation threshold
(\ref{aniso_ration_app}):
\begin{equation}
    \overline{\sigma_{\alpha}} /
    \overline{\sigma_{\beta}} = 1 +
    \frac{z-1}{z} \frac{\sigma_{\alpha}^2-\sigma_{\beta}^2}
  {\sigma_{\alpha}\sigma_{\beta}}\epsilon + \text{O}(\epsilon^2).
\end{equation}
This result is analogous to (\ref{skal_shklovskii}) for the case
when $\alpha$ and $\beta$ are associated with the parallel and the
perpendicular components, respectively. Thus, we receive the
critical exponent, $\lambda=1$, which is consistent with the
exponent obtained by Straley \cite{Straley} by an analogous method.

\begin{figure}
\epsfxsize=3.0in \epsffile{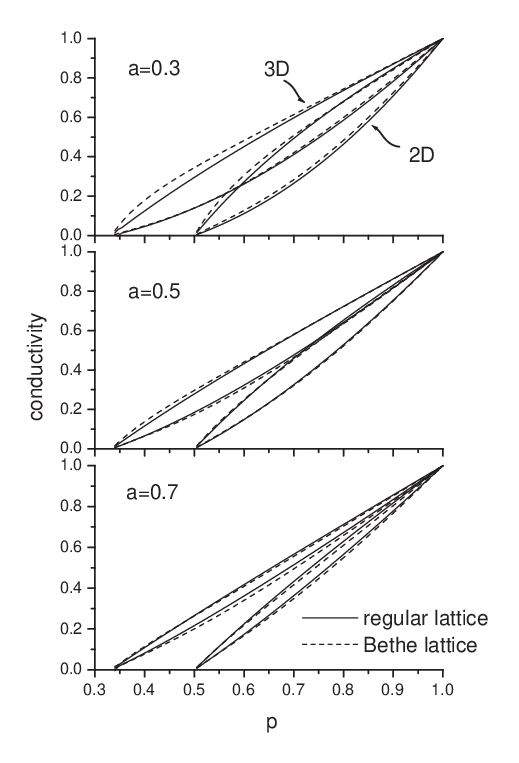} \caption{Comparison of
$\overline{\sigma}_{\text{(I)}}$ given by eq.
(\ref{sigma_final_new}) ($z=3,4$) with the exact solution of
anisotropic RRN on the square and cubic lattice for three magnitudes
of the local anisotropy, $a=\sigma_{\bot}/\sigma_{\|}=0.3$, $0.5$,
and $0.7$. The concave and convex curves represent macroscopic
conductivity in the direction of larger and smaller local
conductivity, $\sigma_{\|}$ and $\sigma_{\bot}$, respectively}
\label{FigBern}
\end{figure}

\section{Comparison with exact solution of Bernasconi}
The topology of the Bethe lattice is quite different from the
regular lattice, but it turns out that both models produce almost
the same values of conductivity normalized to the corresponding
maximum at full occupation. To achieve this correspondence, one uses
the Bethe lattice with z=3 and 4 and the regular lattice of 2D and
3D, respectively. To verify this we compare with the exact solution
of Bernasconi \cite{Bernasconi}. In the two-dimensional case the
exact solution on the square lattice is given by
\begin{equation}
    x = \frac{2}{\pi} \arctan
    \left[\frac{x(p-x)}{a(1-x)(x+p-1)}\right]^{1/2},
\end{equation}
which needs to be solved for $x$ and substituted into
\begin{equation}
    \bar{\sigma}_{\|} = \sigma_{\|}\frac{p-x}{1-x}, \>\>
    \bar{\sigma}_{\bot} = \sigma_{\bot} \frac{x+p-1}{x}
\end{equation}
to obtain the average network conductivity with the conductive
elements $\sigma_{\|,\bot} = \sigma_{\alpha,\beta}$. In the
three-dimensional case of the uniaxial symmetry, Bernasconi provides
the equation
\begin{equation}
    x = 2/\pi\arctan [2U+U^{2}]^{-1/2},
\end{equation}
with
\begin{equation}
        U = \frac{a(1-x)(2p-1+x)}{(1+x)(p-x)},
\end{equation}
which needs to be solved for $x$ and substituted into
\begin{equation}
    \bar{\sigma}_{\|} = \sigma_{\|}\frac{p-x}{1-x}, \>\>
    \bar{\sigma}_{\bot} = \sigma_{\bot} \frac{2p-1+x}{1+x}.
\end{equation}
The latter two equations can be derived in analogy with 2D case
following Bernasconi. We find that both approaches give very close
predictions for a moderate anisotropy in the range of
$a\simeq0.3..1$, see Fig. \ref{FigBern}. The factor $\Delta/p$ in
(\ref{sigma_final}) makes the fit better, especially for the
averaged component corresponding to the preferential conductivity
direction. The deviations start to become significant at higher
values of anisotropy, $a<0.3$. It is clear that the discrepancy is
not due to the finite number of expansion terms over $(z - 1)^{-1}$,
since the truncation of the summation in (\ref{sigma_final}) at
$k=2$, i.e. neglecting the $k=3,4$ terms, preserves the good fit in
the interval [0.3..1] (not shown here). Apparently, the correlations
due to the loops of the regular lattice start to play more and more
pronounced role upon the increase of the intrinsic anisotropy.

In view of quite good conformity of the theory for $z = 3,4$ and the
conductivity on the regular 2D and 3D lattice for moderate
anisotropies, it becomes clear: a) the fact of the fast convergence
of the $(z - 1)^{-1}$ expansion, because the fit becomes better as
the number of expansion terms is increased and b) the fact that the
topology of the Bethe lattice, being quite different from the
regular lattice, is somehow \emph{capable to capture the
correlations of the regular lattice by an adjustment of the
coordination number to a lower integer value}.

\section{Comparison with experimental data}
It is well known that the Kirkpatrick's
\cite{Kirkpatrick} Efective Medium Approximation (EMA),
$\bar{\sigma} \sim (p-p_c)/(1-p_c)$, is the most convenient first
order approximation widely used for experimental data far from the
percolation threshold \cite{Renshaw,Sahimi2}. Also, near the
percolation threshold the empirically observed law is $\bar{\sigma}
\sim (p/p_c-1)^t$, where $t$ is approximately equal to 2 for 3D
systems \cite{Stauffer,Toker,Clerc}. The Bethe lattice theory,
thanks to Stinchcombe \cite{Stinchcombe}, readily explains the
presence of both regimes: $(p-p_c)/(1-p_c)$ and $(p/p_c-1)^2$. It is
therefore not surprising that, in the view its elegance, the Bethe
lattice has been used by us as the central paradigm of the network
modeling.

The discrepancy of the critical exponent of 2 with the value of 3 in
the case of infinite dimensions examined by de Gennes
\cite{deGennes}, we assign to the category of unresolved problems
especially because de Gennes uses the concept of surface which is
never considered in Bethe lattice theories. To make our point clear,
we note that the surfaces may belong to a microscopic or macroscopic
scale in general. Since the infinite branching Cayley tree cannot be
embedded in a finite dimensional space, the macroscopic surfaces in
3D are not the surface sites of the Cayley tree. On the other hand,
the microscopic surfaces could, in principle, be captured by the
Cayley tree, but not by the Bethe lattice where the surface sites
are neglected by definition and $z=const$. Regarding the critical
exponents, Bethe lattice approximation captures only weak
correlations which is usually not sufficient at the critical point,
but the approximation is better than the mean-field \cite{Gujrati}.
The exact mean-field limit is achieved when $z\rightarrow\infty$.
Thus, in general, the Bethe lattice critical exponents are of
specific nature, which implies that they may or may not coincide
with the real values. An example of the coincidence can be found in
the classical Flory-Stockmayer theory of sol-gel transition where
the critical exponents $\sigma$ and $\tau$ are found to be close to
the real values of 3D \cite{StaufferPhysRep}.

\section{Macroscopic vs Microscopic}
This section is devoted to the analysis of the paper by Straley
\cite{Straley}. There one finds the statement: ``...the macroscopic
conductivity is the average current in a link in the presence of a
unit external electric field''. Let us analyze this definition
carefully on the anisotropic Bethe lattice. The local current
through the potential difference between two neighboring nodes is
given by
\begin{equation*}
    I_{\alpha,\beta}^{n} = (V_{\alpha,\beta}^{n-1}-V_{\alpha,\beta}^{n})
    \sigma_{\alpha,\beta},
\end{equation*}
so that the macroscopic conductivity is found from
\begin{equation*}
    \Sigma^{n}_{\alpha,\beta} = (Q_{n}-1)\sigma_{\alpha,\beta},
\end{equation*}
where
\begin{equation*}
    \Sigma^{n}_{\alpha,\beta} =
    \frac{I_{\alpha,\beta}^{n}}{V_{\alpha,\beta}^{n}}, \>\>\> \text{and}
    \>\>\>     Q_{n}=\frac{V_{\alpha}^{n-1}}{V_{\alpha}^{n}} =
    \frac{V_{\beta}^{n-1}}{V_{\beta}^{n}}.
\end{equation*}
The fact that $Q_{n}$ is independent of $\alpha$ and $\beta$ follows
from the occupation probability $p$ being independent of those
indices. For simplicity we show the proof only for the case of fully
occupied, $p=1$, Bethe lattice. With the help of the Kirchoff's law
which states that the sum of the currents on each internal site is
zero,
\begin{equation*}
    V_{i} = \frac{\sum_{j} \sigma_{ij}V_{j}}{\sum_{ij}\sigma_{ij}},
\end{equation*}
the formulation of the problem in terms of the recurrent relations
is straightforward. For instance, for the case of coordination
number $z=4$ we have the following recursive relations
\begin{eqnarray*}
    V_{\alpha}^{n} &=& \frac{V_{\alpha}^{n+1}\sigma_{\alpha} +
    2V_{\beta}^{n+1}\sigma_{\beta} +
    V_{\alpha}^{n-1}\sigma_{\alpha}}{2\sigma_{\alpha}+2\sigma_{\beta}}, \\
    V_{\beta}^{n} &=& \frac{V_{\beta}^{n+1}\sigma_{\beta} +
    2V_{\alpha}^{n+1}\sigma_{\alpha} +
    V_{\beta}^{n-1}\sigma_{\beta}}{2\sigma_{\alpha}+2\sigma_{\beta}}.
\end{eqnarray*}
Dividing both sides of the equations by $V_{\alpha}^{n-1}$ and
$V_{\beta}^{n-1}$, respectively, gives the recursive relations for
the ratios
$Q_{\alpha,\beta}^{n}=V_{\alpha,\beta}^{n+1}/V_{\alpha,\beta}^{n}$
and
$Y_{\alpha,\beta}^{n}=V_{\alpha,\beta}^{n+1}/V_{\beta,\alpha}^{n}$.
Performing the iterations from an arbitrary initial values of the
ratios, one finds $Q_{\alpha,\beta}^{n}$ being independent of
$\alpha$ and $\beta$, $Q_{\alpha,\beta}^{n}=Q_{n}$. In the limit of
very large number of iterations, one arrives to the fix-point
$Q_{n}\rightarrow Q$ with the conductivity expressed as
\begin{equation*}
    \Sigma_{\alpha,\beta} = (Q-1) \sigma_{\alpha,\beta},
\end{equation*}
which tells that the ratio of the conductivities in two directions
is just the ratio of the conductive elements
\begin{equation*}
    \Sigma_{\alpha}/\Sigma_{\beta} = \sigma_{\alpha}/\sigma_{\beta}.
\end{equation*}
Here, the factor $Q-1$ measures the average ratio of potentials of
two neighboring nodes which gives the average current in a link. We
note this is the maximum value anisotropy as a function of $p$
assuming that the resistors are distributed homogeneously. While
this relationship is the correct one for the regular lattice, this
model fails to explain considerably higher values of macroscopic
anisotropy of the Earth mantle as compared to the microscopic ones.
Presuming that the distribution of conductive inclusions is
homogeneous, only a fractal structure could possibly explain this
experimental observation. Thus, the definition of the macroscopic
conductivity on Bethe lattice proposed by Straley seems to be
incapable of accounting for the anisotropy growth upon the change
from the microscopic to the macroscopic scale.

\section{Discussion and conclusions}
We propose the model of the resistor network that has a property of
anisotropy in a sense that the conductivities of the resistors
differ with respect to the lattice bond type. We solve the problem
in the framework of the anisotropic Bethe lattice approximation. The
mathematical problem is formulated in terms of a nonlinear integral
equation, which is solved asymptotically using series expansions in
two limiting cases: near the percolation threshold  and near the
mean-field limit of $z\rightarrow\infty$.

It seems that the Bethe lattice may be a suitable model for the
conductivity anisotropy of geological resistor networks far from the
percolation threshold. Generally speaking, a Bethe lattice branch,
see Fig. \ref{FigBetheLettice}, is one of many possible models of a
statistically homogeneous random graph. By homogeneity we mean
allowing for only very small fluctuations of co-ordination numbers
of the nodes. For the purpose of the large-scale characterization of
the network, the co-ordination numbers of different nodes (vertices)
can be approximately considered as uniform and equal to an average
value. One possibility is to use the wholly tree-like structure in
which the average shortest path length scales as a power of the
total number of vertices \cite{Burda}. Another possibility would be
to use the model of small-world network where the average shortest
path is signified by the logarithmic dependence on the graph size
\cite{Bolobas}. These are two theoretical examples of the different
specific classes of real-world networks empirically observed. The
significant anisotropies observed in geophysics at the macroscale
could be explained by the formation of fractal structures in a
microscale. In the present model, the macroscopic observable
anisotropy is the property of entire network and the local
(intrinsic) anisotropy is associated with the anisotropy in
conductivity at a branching point of the Bethe lattice. The former
is defined as $\overline{a} = \overline{\sigma_{\beta}} /
\overline{\sigma_{\alpha}}$, whereas the latter is essentially the
ratio $a = \sigma_{\beta} / \sigma_{\alpha}$ being the \emph{only}
parameter entering eqs.
(\ref{sigma_final},\ref{Stinch},\ref{sigma_final_new}). We find that
the present theory is capable of producing the strong global
anisotropy, $\overline{a}$, at small local anisotropy, $a$, in the
case when $z$ is large and $p \gg p_{c}$. Indeed, in this limit the
conductivity is given by the mean-field formula:
$\overline{\sigma_{\alpha}}_{\text{(I)}} \approx zp
    \sigma_{\alpha}^{2}/(\sigma_{\alpha} + \sigma_{\beta})$,
which yields $\overline{a}_{\text{(I)}} \approx a^{2}$. In many
cases the dynamical networks are driven to criticality, but the
networks driven far away from critical point are also realizable in
principle and possible to occur in nature. Interestingly, that the
previous theories based on anisotropic occupation probability
\cite{Blane,Gavrilenko} predicted the opposite: at strong local
anisotropy - weak global one.

In course of our derivation we employed the approximation that the
number of the special directions $n$ remains finite as $z
\rightarrow \infty$. Although, it is not possible to give a simple
geometrical picture relating $n$ to some normal space coordinates,
since Bethe lattice cannot be embedded in a finite dimensional
space, the number $n$ seems to be associated with the the number of
possible directions of the spanning cluster near the percolation
threshold. The problem needs to be resolved on more rigorous
topological grounds.

\section{Acknowledgments}
FS is grateful to Professor P.D. Gujrati for many fruitful
discussions and kind attention. We also thank the anonymous referees
for their valuable comments and constructive criticism. The
financial support of the German Federal Ministry of Education and
Research (BMBF) under the project CarboNet No. 03X0504E is
gratefully acknowledged.

\appendix
\numberwithin{equation}{section}
\section{Anisotropic Bethe lattice theory}
This section is essentially the anisotropic generalization of Ref.
\cite{Stinchcombe} with the intermediate steps shown explicitly in
Ref. \cite{Hughes}. For a branch starting from an $\alpha$-bond and
its $z-1$ next generation subbranches (see Fig.
\ref{FigBetheLettice}), the set of conductivities is defined:
\begin{equation*}
    \{ b_{\alpha}^{(i)} \} = b_{\alpha}^{(0)},
    b_{\alpha}^{(1)} ,...,
b_{\alpha}^{(n_{\alpha}-1)}, b_{\beta}^{(n_{\alpha})}, ...,
b_{\beta}^{(z-1)},
\end{equation*}
which are zero or finite according as the corresponding root bonds
are empty or occupied, where the index $i = 0$ is reserved for the
branch origin. These subbranches are connected in parallel, so that
the conductivities $b_{\alpha}^{(0)}$ are given by
\begin{equation}
  b_{\alpha}^{(0)} = \sum_{i=1}^{n_{\alpha}-1}
  \frac{\sigma_{\alpha}
  b_{\alpha}^{(i)}}{\sigma_{\alpha}+b_{\alpha}^{(i)}} +
  \sum_{i=n_{\alpha}}^{z-1} \frac{\sigma_{\beta} b_{\beta}^{(i)}}
  {\sigma_{\beta}+b_{\beta}^{(i)}}
\end{equation}
Both $b_{\alpha}^{(i)}$ ($b_{\beta}^{(i)}$) and $\sigma_{\alpha}$
($\sigma_{\beta}$), which are the branch and bond conductivities,
respectively, are random variables distributed with some probability
density functions. In order to determine the branch distribution
functions $\phi_{\alpha}(b^{(0)})$ defined in (\ref{phi_norm_cond})
and (\ref{b_av_def_1}) we average over various resistor
configurations on the lattice using the distribution functions
$\phi_{\alpha}(b^{(i)})$ and $g_{\alpha}(\sigma^{(i)})$ defined for
the conductivities of subbranches and individual bonds,
respectively, so that
\begin{equation}
    g_{\alpha}(\sigma) = p\,\delta(\sigma-\sigma_{\alpha}) +
    (1-p)\delta(\sigma). \label{g_alpha_def}
\end{equation}
Being more specific we determine $\phi_{\alpha}(b)$ [note that the
superscript (0) is suppressed for brevity] by performing an
asymptotic analysis of
\begin{eqnarray}
  \phi_{\alpha}(b) &=&  \prod_{i=1}^{n_{\alpha}-1} \left( \int_{0}^{\infty}
    d\sigma^{(i)}\, g_{\alpha}(\sigma^{(i)}) \int_{0}^{\infty} db^{(i)} \,
    \phi_{\alpha}(b^{(i)}) \right)\times  \nonumber \\
   && \prod_{i=n_{\alpha}}^{z-1} \left( \int_{0}^{\infty} d\sigma^{(i)}\,
    g_{\beta}(\sigma^{(i)}) \int_{0}^{\infty} db^{(i)}\, \phi_{\beta}(b^{(i)})
    \right) \times \nonumber \\
    && \delta\left( b - b_{\alpha}^{(0)} \right), \label{phi}
\end{eqnarray}
Since $\phi_{\alpha}(b)$ is actually a series of delta functions, it
is convenient to introduce the Laplace transform of
$\phi_{\alpha}(b)$, generally known as the moment-generating
function
\begin{equation}
    B_{\alpha}(q) \equiv \int_{0}^{\infty} e^{-qb} \phi_{\alpha}(b)
    db. \label{B_alpha_def}
\end{equation}
In the present study, this quantity is named the branch generating
function. Equation (\ref{B_alpha_def}) combined together with
(\ref{b_av_def_1}) leads to
\begin{equation}
    \bar{b}_{\alpha} = -B_{\alpha}^{'}(0), \label{b_av_def}
\end{equation}
which means the average branch conductivity is just the negative
first derivative of $B_{\alpha}(q)$ evaluated at $q=0$. Taking the
Laplace transform of eq.(\ref{phi}) it can be shown that
\begin{equation} \label{B_C_alpha_C_beta}
    B_{\alpha}(q) =
    C_{\alpha}(q)^{n_{\alpha}-1}C_{\beta}(q)^{n_{\beta}},
\end{equation}
where
\begin{eqnarray}
  \lefteqn { C_{\alpha}(q) =
  }  \nonumber \\
  && \int_{0}^{\infty} d\sigma\, g_{\alpha}(\sigma)
    \int_{0}^{\infty}  db\, \phi_{\alpha}(b)
    \exp \left( -\frac{q\sigma b}{\sigma + b} \right).
    \label{C_def}
\end{eqnarray}
Therefore, on account of (\ref{B_C_alpha_C_beta}), $C_{\alpha}(q)$
and $C_{\beta}(q)$ can be named the subbranch generating functions
and, in analogy with (\ref{b_av_def}), one can define the subbranch
average conductivity as
\begin{equation}
    \bar{b}_{\alpha}^{(i)} = -C^{\,'}_{\alpha}(0).
    \label{bi_av_def}
\end{equation}
Since $\phi_{\alpha}(b)$ and $g_{\alpha }(\sigma )$ are the
probability densities normalized to unity, see (\ref{phi_norm_cond})
and (\ref{g_alpha_def}), respectively, the boundary condition for
$C_{\alpha}(q)$ at $q=0$ is
\begin{equation}
    C_{\alpha}(0) = 1. \label{bound_cond_0}
\end{equation}
The other boundary condition at $q=\infty$ is identified as the
probability to have the finite cluster, $R$, since the main
contribution to the integral (\ref{C_def}) comes from the
neighborhood of  $b=0$ or $\sigma = 0$:
\begin{equation}
    C_{\alpha}(\infty) = R. \label{bound_cond_inty}
\end{equation}
After some algebra \cite{Stinchcombe} which involves an additional
Laplace transform that introduces a new variable $t$, one obtains
the integral equation
\begin{eqnarray}
  \lefteqn { \int_{0}^{\infty} e^{-tq} C_{\alpha}(q) dq =
  \int_{0}^{\infty} d\sigma g_{\alpha}(\sigma) (t+\sigma)^{-1} \times}
  \nonumber \\
  && \left[ 1 +\frac{\sigma^2}{t+\sigma}
    \int_{0}^{\infty} \exp \left( -\frac{q\sigma t}{\sigma+t}
    \right)  B_{\alpha}(q) dq \right] \label{start_eq}
\end{eqnarray}
which is the final exact result to be solved asymptotically.

\section{Near-mean-field expansion}
Consider the integrals on both sides of eq. (\ref{start_eq}). These
integrals will be approximated for large $t$ values using the
Laplace method \cite{Bender}. The method is based on the idea that
the main contribution to the integrals comes from the neighborhood
of $q=0$, which makes it possible to use the Taylor series expansion
as follows
\begin{eqnarray}
    \lefteqn {C_{\alpha}(q)=e^{\ln \left[C_{\alpha}(0)+qC_{\alpha}^{\,'}(0)+
    ...\right]}} \nonumber \\
    && =e^{qC^{\,'}_{\alpha}(0)}\left[ 1 + \sum_{l=2}^{\infty}
    a_{\alpha}^{(l)} q^{l} \right],\>\>\> \text{for} \>\>\> q\ll1.
    \label{C_expans}
\end{eqnarray}
This defined the coefficients $a_\alpha^{(l)}$. In addition, we have
the $d_\alpha^{(l)}$ coefficients given by
\begin{equation}
  {C_{\alpha}( q )}^{m}  = {e^{mqC^{\, '}_{\alpha} \left( 0 \right) }}
\sum _{l=0}^{\infty } d_{\alpha}^{(l)} {q}^{l}, \label{Cm_expans}
\end{equation}
where $d_{\alpha}^{(0)}=1$. Substituting (\ref{Cm_expans}) and its
conjugate $\beta$ analog into (\ref{B_C_alpha_C_beta}) and then
using this in (\ref{start_eq}), one obtains
\begin{eqnarray}
    \lefteqn { \int_{0}^{\infty} e^{-tq} C_{\alpha}(q)dq =
    \int_{0}^{\infty} du g_{\alpha}(u) } \nonumber \\
    && \left\{ \frac{1}{t-\tau_{\alpha}(u)}+\frac{u^2}{(t+u)^2}
    \sum_{k=2}^{\infty}d_{\alpha}^{(k)}
    \frac{k!}{s^{k+1}_{\alpha}} \right\}, \label{App_fist_st}
\end{eqnarray}
where
\begin{eqnarray}
  s_{\alpha} &=& \frac{ut}{u+t} - (n_{\alpha}-1) C^{\,'}_{\alpha} (0) -
  n_{\beta} C^{\,'}_{\beta} (0), \>\>\; \label{s_alpha} \\
    \tau_{\alpha}(u) &=& \frac{u\left[(n_{\alpha}-1)C^{\,'}_{\alpha}(0)+
    n_{\beta}C^{\,'}_{\beta}(0)\right]}
    {u-(n_{\alpha}-1)C^{\,'}_{\alpha}(0)-n_{\beta}C^{\,'}_{\beta}(0)}.
    \label{tau}
\end{eqnarray}
The formula (\ref{App_fist_st}) represents an expansion in inverse
powers of $z$ which is seen from (\ref{s_alpha},\ref{tau}).
Inversion of the Laplace transform in (\ref{App_fist_st}) yields
\begin{eqnarray*}
    \lefteqn { C_{\alpha}(q) = \int_{0}^{\infty} du \,g_{\alpha}(u)\,
    \exp \left[q \tau_{\alpha}(u) \right] \times} \\
    && \left[ 1+\sum_{k=2}^{\infty} d_{\alpha}^{(k)} k! \sum_{r=0}^{k-1}
    \frac{_{k-1} C_{r}}{(k-r)!} \times \right.\\
    && \left. \frac{u^{2(k-r)}q^{k-r}}{\left[u-(n_{\alpha}-1)
    C^{\,'}_{\alpha}(0)-n_{\beta}C^{\,'}_{\beta}(0)
    \right]^{2k-r}}\right],
\end{eqnarray*}
where $_{k-1} C_{r}$ are the binomial coefficients. The equation
right above is combined with (\ref{C_expans}), then one equates term
by term the factors of the successive powers of $q$, and obtains
\begin{equation}
    1=\int_{0}^{\infty} du g_{\alpha}(u)
\end{equation}
for $l=0$ and
\begin{eqnarray}
    0 &=& \int_{0}^{\infty} du g_{\alpha}(u)[\tau_{\alpha}(u)-C^{\,'}_{\alpha}(0)]
    \nonumber \\
    && +\sum_{m=2}^{\infty}d_{\alpha}^{(m)} I^{(m10)}_{\alpha}
    \label{mean_expans} \\
    a_{\alpha}^{(k)} &=& a_{\alpha}^{(k)0} + \sum_{m=2}^{\infty}
    d_{\alpha}^{(m)} \times  \nonumber \\
    && \sum_{s=0}^{\text{min}\{m-1,\,k-1\}} I_{\alpha}^{(mks)}, \>\>
    k \geq 2, \label{a_expans}
\end{eqnarray}
for $l=1$, where
\begin{equation}
    a_{\alpha}^{(k)0} = \int_0^{\infty} du\,g_{\alpha}(u)
    \frac{\left[\tau_{\alpha}(u)-C^{\,'}_{\alpha}(0)\right]^k}{k!}
    \label{a0}
\end{equation}
and
\begin{eqnarray}
    \lefteqn { I_{\alpha}^{(mks)} = \frac{m!\,
    _{m-1}C_{s}}{(s+1)!\left[ k-(s+1)
    \right]!} \int_{0}^{\infty} du g_{\alpha}(u) \times } \nonumber \\
    && \frac{u^{2(s+1)}[\tau_{\alpha}(u)-
    C^{\,'}_{\alpha}(0)]^{k-(s+1)}}{[u-(n_{\alpha}-1)C^{\,'}_{\alpha}(0)
    -n_{\beta}C^{\,'}_{\beta}(0)]^{m+s+1}}. \label{Imks}
\end{eqnarray}
The first term on the right-hand-side of (\ref{mean_expans}) is the
representation of the mean-field limit $z\rightarrow\infty$,
\begin{equation} \label{C_mean}
    0=\int_{0}^{\infty} du
    g_{\alpha}(u)[\tau_{\alpha}(u)-\overline{C}^{\,'}_{\alpha}(0)],
\end{equation}
and the sum over $m$ gives the corrections in inverse powers of $z$.
Two equations, obtained by the interchange of $\alpha$ and $\beta$
in (\ref{C_mean}), will be solved neglecting $n_{\alpha} = n$ as it
is a constant negligibly small compared to $z-1$. [Here, in order to
keep up with the Stinchcome's results, we expand in powers of
inverse $z-1$ and not $z$, which is equivalent]. Solving
(\ref{C_mean}) we get, as the first solution, the isotropic
mean-field conductivity:
\begin{equation}
    \overline{C}^{\,'}_{\alpha}(0)_{\text{iso}} = - \sigma_{\alpha}
    (p-p_{c}). \label{C_mean_sol_iso}
\end{equation}
Additionally, we obtain
\begin{equation}
    \overline{C}^{\,'}_{\alpha}(0) = -\frac{\sigma_{\alpha}
    \sigma_{\beta} (p-p_{c})p}
    {\sigma_{\alpha}p_{c} + \sigma_{\beta} (p-p_{c})},
    \label{C_mean_sol}
\end{equation}
which is the anisotropic solution of main interest for us.

We now move to some elaboration regarding the orders of correction
contained in (\ref{mean_expans})-(\ref{Imks}). One finds that
$I_{\alpha}^{(210)} d_{\alpha}^{(2)}$ is of the order $(z-1)^{-2}$,
since $I_{\alpha}^{(m10)}$ and $d_{\alpha}^{(m)}$ are of the orders
$(z-1)^{-(m+1)}$ and $(z-1)^{m/2}$, respectively. Note that the
correction to $a_{\alpha}^{(2)0}$ given by the first term of the sum
in (\ref{a_expans}) affects $I_{\alpha}^{(210)} d_{\alpha}^{(2)}$ by
 $(z-1)^{-4}$ order. Thus, to have the final result up to and
including $\text{O}([z-1]^{-4})$, the first term in the sum given by
(\ref{a_expans}) should be taken into account, but only
$a^{(m)0}_{\alpha}$ can be used for $m>2$. In addition, when
approximating $a_{\alpha}^{(m)}$ and $I_{\alpha}^{(m10)}$ with
$m>2$, we use the replacement $C^{\,'}_{\alpha}(0) =
\overline{C}^{\,'}_{\alpha}(0)$. This is perfectly acceptable if
$m>2$, since any correction to this would be of $(z-1)^{-2}$ order,
and hence would contribute to $a^{(m)0}_{\alpha}$ as $(z-1)^{-2m}$.
To compute the error introduced by this substitution for $m=2$, we
expand $T_{\alpha} = I_{\alpha}^{(210)} a^{(2)0}_{\alpha}$ near
$\bar{T}_{\alpha} = \overline{I}_{\alpha}^{(210)}
\overline{a}^{(2)0}_{\alpha}$:
\begin{equation} \label{T_exp}
    T_{\alpha} = \bar{T}_{\alpha} + (D^{\alpha} T_{\alpha})
    \Delta C_{\alpha}+(D^{\beta}T_{\alpha})\Delta C_{\beta},
\end{equation}
where $\Delta C_{\alpha} = C^{\,'}_{\alpha}(0) -
\overline{C}^{\,'}_{\alpha}(0)$, $D^{\beta}T_{\alpha} = [\partial
T_{\alpha} /
\partial C^{\,'}_{\beta}(0)] _{\overline{C}^{\,'}_{\beta}(0)}$.
Additionally, from (\ref{mean_expans}) with the term $m=2$ only, one
has
\begin{equation} \label{T_bar_exp}
    -\bar{T}_{\alpha} = (D^{\alpha}A_{\alpha}) \Delta C_{\alpha}
    + (D^{\beta}A_{\beta}) \Delta C_{\beta}
\end{equation}
where $D^{\beta}A_{\alpha} = [\partial \int du g_{\alpha}(u)
[\tau_{\alpha} - C^{\,'}_{\alpha}(0)] /
\partial C^{\,'}_{\beta}(0)] _{\overline{C}^{\,'}_{\beta}(0)}$. The
computation of the coefficients yields
\begin{eqnarray}
    D^{\alpha}T_{\alpha} &=& -1 + \text{O}([z-1]^{-1}), \nonumber \\
    D^{\beta}T_{\alpha} &=& 0, \nonumber \\
    D^{\alpha} A_{\alpha} &=& 0, \nonumber \\
    D^{\beta}A_{\alpha} &=& (z-1) \overline{I}^{(310)}_{\alpha}
    \overline{d}^{(2)0}_{\alpha}. \label{part_derivs}
\end{eqnarray}
Equations (\ref{T_exp})-(\ref{part_derivs}) combined together give
\begin{eqnarray}
    T_{\alpha} &=& \bar{T}_{\alpha}\left(1 +
    (z-1) \overline{I}^{(310)}_{\alpha}
    \overline{d}^{(2)0}_{\alpha} \frac{\bar{T}_{\beta}}
    {\bar{T}_{\alpha}}\right) \nonumber \\
    &&= \bar{T}_{\alpha}+ \text{O}([z-1]^{-4}).
    \label{T_exp_fin}
\end{eqnarray}
Substituting this into (\ref{mean_expans}), one obtains the
expression that contains all corrections up to $(z-1)^{-4}$ order:
\begin{eqnarray}
  \lefteqn {0 = \int g_{\alpha}(u)\left[ \tau_{\alpha}(u)-C^{\,'}_{\alpha}(0) \right]
        + (z-1)a^{(2)0}_{\alpha} I_{\alpha}^{(210)} \times } \nonumber \\
    && \left\{ 1+(z-1)
        I_{\alpha}^{(220)} + (z-1)^2 a^{(2)0}_{\alpha} I_{\alpha}^{(310)}
    \frac{I_{\beta}^{(210)}}
        {I_{\alpha}^{(210)}} \right\}  \nonumber \\
    &&+ I_{\alpha}^{(310)} \,d^{(3)0}_{\alpha}
    + I_{\alpha}^{(410)} \,d^{(4)0}_{\alpha}+
    I_{\alpha}^{(510)}\,d^{(5)0}_{\alpha} \nonumber \\
    && + I_{\alpha}^{(610)} \,d^{(6)0}_{\alpha}.
    \label{final_exp_3}
\end{eqnarray}

Direct computation of the eqs. (\ref{a0}) and (\ref{Imks}) yields
\begin{eqnarray}
  \overline{a}^{(m)0}_{\alpha} &=& \frac{1}{m!}(-1)^m
  \overline{C}^{\,'}_{\alpha}(0)^{m}(1-p)\times  \nonumber \\
  && \left[1+(-1)^{m}\left(\frac{1-p}{p}\right)^{m-1} \right], \label{a_m_comp}\\
  \overline{I}_{\alpha}^{(m10)} &=& \frac{m!\left[\overline{C}^{\,'}_{\alpha}(0)+
                        \sigma_{\alpha}\,p\right]^{m+1}}{\sigma^{m-1}_{\alpha}\,p^{\,m}}
                        \label{Im10_comp}\\
  \overline{I}_{\alpha}^{(220)} &=& \frac{2(1-p)\overline{C}^{\,'}_{\alpha}(0)
                      \left[\overline{C}^{\,'}_{\alpha}(0)+\sigma_{\alpha}\,p\right]^{3}}
                      {\sigma_{\alpha}\,p^{\,2}} \label{I220_comp}
\end{eqnarray}
Finally, substituting (\ref{a_m_comp})-(\ref{I220_comp}) into
(\ref{final_exp_3}) and using (\ref{b_av_def}), we get the
anisotropic conductivity in the form of a series expansion in
successive powers of the inverse coordination number:
\begin{equation}
    \overline{b_{\alpha}}_{\text{(I)}} = -\sigma_{\alpha}
    \left( -\frac{p \sigma_{\beta} \Delta }
    {\sigma_{\alpha}p_{c}+\sigma_{\beta} \Delta} +
    \frac{\Delta}{p}
    \sum_{n=2}^{\infty}G^{(n)} \right), \label{sigma_final_App}
\end{equation}
where
\begin{eqnarray}
  G^{(2)} &=& \frac{p_{c}^{2}}{p^2}\,\Delta \,s, \nonumber \\
  G^{(3)} &=& \frac{p_{c}^3}{p^5}\,\Delta^2 s [ p(2p-1)+3\Delta s
  ], \nonumber \\
  G^{(4)} &=& \frac{p_{c}^{4}}{p^6}\,\Delta^{2}s\left[
                    3s^{2}\Delta - 2sp + \Delta (1-3p+3p^{2})
                    \textcolor[rgb]{1.00,1.00,1.00}
                    {\frac{\frac{}{}}{\frac{}{}}}
                    \right. \nonumber \\
  &&\left.+\, 10\Delta^{2}s\left( \frac{2p-1}{p} \right) +
                    15\Delta^{3}s^2 \frac{1}{p^2} \right], \nonumber \\
  G^{(n)} &=& \text{O}(p_{c}^{n}), \nonumber \\
  \Delta &=& p-p_c, \>\>\>\> \text{and} \>\>\>\> s = 1-p. \label{StinchA}
\end{eqnarray}
The expansion (\ref{StinchA}) coincides with the result of Ref.
\cite{Stinchcombe} except that factor 5 in front of
$\Delta^{2}s\left( \frac{2p-1}{p} \right)$ for $G_{\alpha}^{(4)}$
needs to be replaced by 10 according to us.

\section{Investigation of critical indices}
In this appendix we investigate the critical exponents by means of
an asymptotic analysis of the integral equation (\ref{start_eq}) for
$p$ approaching $p_{c}$ from above introducing a small parameter
\begin{equation}
  \epsilon = \frac{p-p_{c}}{p_{c}}. \label{eps_deff}
\end{equation}

On the one hand, it is known \cite{Stauffer} that the percolation
probability $P$, defined on the Bethe lattice as $P=1-R^{z}$, can be
expanded near the percolation threshold in series:
\begin{equation*}
    P(\epsilon) = B \epsilon + C \epsilon^2 + \text{O}(\epsilon^3),
\end{equation*}
where $B$ and $C$ are constants and hence
\begin{equation}
    R = 1 - \delta^{(1)} \epsilon - \delta^{(2)} \epsilon^2 +
        \text{O}(\epsilon^3). \label{R_near_pc}
\end{equation}
Here,
\begin{equation}\label{delta_1}
    \delta^{(1)}=2/(z-2),
\end{equation}
while the numerical value of $\delta^{(2)}$ has no significance for
us, as shown in the analysis set forth below.

On the other hand, the anisotropic conductivity expansion in terms
of $\epsilon$ has not previously been addressed. Motivated by the
analysis of Ref. \cite{Stinchcombe}, we propose a trial solution to
(\ref{start_eq}) of the form
\begin{equation}
    C_{\alpha}(q) = R + \epsilon C_{\alpha}^{(1)}(q) +
    \epsilon^2 C_{\alpha}^{(2)}(q),
    \label{C_near_pc}
\end{equation}
where $C_{\alpha}^{(1,2)}(q)$ are slowly varying functions of $q$
described by the scaling relations
\begin{equation}
    C_{\alpha}^{(1,2)}(q) = f_{\alpha}^{(1,2)}(c_{\alpha} \epsilon q),
    \label{C_slow}
\end{equation}
with $c_{\alpha}$ being a constant to be determined later.

Now, the variables $s = t/(c_{\alpha} \epsilon)$ and $y = c_{\alpha}
\epsilon q$ are defined. Substituting (\ref{R_near_pc}) and
(\ref{C_near_pc}) into the left-hand-side of (\ref{start_eq}),
multiplying both sides by $t$ and expressing the integrals in terms
of the new variables, one finds
\begin{eqnarray}
  \lefteqn { \int _{0}^{\infty }
  \!{e^{-sy}} [C_{\alpha}(y)-1]  {dy} = p_{c}\frac{\sigma_{\alpha}
  (1+\epsilon)} {(\sigma_{\alpha} + c_{\alpha} \epsilon s)^2} \times} \nonumber \\
   & &  \int _{0} ^{\infty }\!{ \exp \left( {-{\frac
{sy\sigma_{\alpha}}{\sigma_{\alpha} + c_{\alpha} \epsilon s}}}
\right) } \left\{C_{\alpha}^{n_{\alpha}-1}
C_{\beta}^{n_{\beta}}-1\right\}{dy}, \label{Int_eq_near_pc_1}
\end{eqnarray}
where $C_{\alpha}(y)$ are given by (\ref{C_near_pc}) and
(\ref{C_slow}). A set of equations is obtained equating the
$\epsilon$-expansion coefficients of the same order in the left- and
right-hand-side of the integral equation (\ref{Int_eq_near_pc_1}).
Firstly, equating the terms linear in $\epsilon$ we obtain
\begin{equation}
    f^{(1)}_{\alpha}(y) = p_{c} [(n_{\alpha}-1) f^{(1)}_{\alpha}(y)
    + n_{\beta} f^{(1)}_{\beta}(y)]. \label{f1_1}
\end{equation}
As a result, the first correction is isotropic,
\begin{equation}
    f^{(1)}_{\alpha}(y) = f^{(1)}_{\beta}(y) \equiv f^{(1)}(y). \label{f1_2}
\end{equation}
Secondly, equating the terms proportional to $\epsilon^2$ and using
(\ref{f1_2}) yields
\begin{eqnarray}
  \lefteqn { \int_{0}^{\infty} e^{-sy} f^{(2)}_{\alpha}(y) dy =
  \int_{0}^{\infty} dy e^{-sy} \times } \nonumber \\
  &&\left\{ \frac{1}{\delta^{(1)}}
    (f^{(1)}-\delta^{(1)})^2 \right. +
    (f^{(1)}-\delta^{(1)}) \times  \nonumber \\
  && \left. \left[1 + \frac{c_{\alpha}}{\sigma_{\alpha}}
    (-2s+s^{2}y) \right]+ p_{c} [(n_{\alpha}-1) f^{(2)}_{\alpha}
    + n_{\beta} f^{(2)}_{\beta}] \right\}.\nonumber \\ \label{f2_1}
\end{eqnarray}

\emph{Isotropic solution:} When substituting
$c_{\alpha}=\sigma_{\alpha}$, we recover the isotropic solution $
f^{(2)}_{\alpha} = f^{(2)}_{\beta} = f^{(2)}$ satisfying
\begin{eqnarray*}
  \lefteqn { \int_{0}^{\infty} e^{-sy} f^{(2)}(y) dy =
  \int_{0}^{\infty} dy e^{-sy} \times }  \\
  &&  \left\{ \frac{1}{\delta^{(1)}}
    (f^{(1)}-\delta^{(1)})^2 +
    (f^{(1)}-\delta^{(1)}) \left[1
    -2s+s^{2}y \right] \right.  \\
  && \left. + p_{c} [(n_{\alpha}-1) f^{(2)}(y)
    + n_{\beta} f^{(2)}(y)] \right\} .
\end{eqnarray*}
and
\begin{eqnarray}
    0  &=& \int_{0}^{\infty} dy e^{-sy} \left\{ \frac{1}{\delta^{(1)}}
    (f^{(1)}-\delta^{(1)})^2 \right.  \nonumber \\
    && \left.+ (f^{(1)}-\delta^{(1)}) \left[1-2s+s^{2}y \right]
    \right\}  \label{f1_eq}
\end{eqnarray}
This gives a simple solution:
\begin{equation}
    f^{(1)}(y)_{\text{iso}} = \delta^{(1)} \xi(y), \label{f1_xi}
\end{equation}
where $\xi$ is determined by solving numerically the differential
equation of the second order,
\begin{equation}
    y \xi'' = \xi(1-\xi), \>\>\> \xi(0)=1, \>\>\> \xi(\infty)=0,
    \label{xi}
\end{equation}
which gives $\xi'(0)=-0.762$.

\emph{Anisotropic solution:} Here, we try $c_{\alpha} \neq
\sigma_{\alpha}$ in (\ref{f2_1}) to obtain another solution. To
simplify (\ref{f2_1}), we combine it with (\ref{f1_eq}), which
yields
\begin{eqnarray*}
  \lefteqn { \int_{0}^{\infty} e^{-sy} f^{(2)}_{\alpha}(y) dy =
  }  \\
  && \int_{0}^{\infty} e^{-sy} \left\{
    (f^{(1)}-\delta^{(1)}) \left[\frac{c_{\alpha}}
    {\sigma_{\alpha}}
    -1 \right](-2s+s^{2}y)  \right.  \\
  && \left. + p_{c} [(n_{\alpha}-1) f^{(2)}_{\alpha}(y)
    + n_{\beta} f^{(2)}_{\beta}(y)] \right\} dy.
\end{eqnarray*}
From this equation follows that
\begin{equation}
    f^{(2)}_{\alpha} - p_{c} [(n_{\alpha}-1) f^{(2)}_{\alpha}
    + n_{\beta} f^{(2)}_{\beta}] =
    \frac{c_{\alpha} - \sigma_{\alpha}}{\sigma_{\alpha}}\delta^{(1)}
    x \xi''(x),   \label{f_2_aniso}
\end{equation}
where $x$ is an arbitrary variable. The quantities
$f^{(2)}_{\alpha}$(x) and $f^{(2)}_{\beta}$(x) have to be symmetric
with respect to the $\alpha \leftrightarrow \beta$ interchange. This
condition is satisfied only in the two following cases:
\begin{eqnarray}
    n_{\alpha} \sigma_{\beta} &=& n_{\beta} \sigma_{\alpha}, \label{n_sigma} \\
    c_{\alpha} &=& \sigma_{\beta}, \label{c}
\end{eqnarray}
or
\begin{eqnarray}
    n_{\alpha} &=& n_{\beta}, \label{n_sigma1} \\
    c_{\alpha} &=& 2\sigma_{\alpha}\sigma_{\beta}/
    (\sigma_{\alpha} + \sigma_{\beta}). \label{c1}
\end{eqnarray}
The second set of conditions, eqs. (\ref{n_sigma1}) and (\ref{c1}),
is inappropriate here due to the condition $z \gg n$ utilized in the
Appendix B. Hence, from (\ref{n_sigma}) we have
\begin{eqnarray}
    n_{\alpha} =
    z\sigma_{\alpha}/(\sigma_{\alpha}+\sigma_{\beta}), \>\>\>
    n_{\beta} = z - n_{\alpha}. \label{n_alpha_n_beta}
\end{eqnarray}
Substitution of (\ref{n_alpha_n_beta}) into (\ref{f_2_aniso}) yields
\begin{equation}
    f^{(2)}_{\alpha}(x) - f^{(2)}_{\beta}(x) = -\delta^{(1)} x \xi''(x)
    \frac{z-1}{z} \frac{\sigma_{\alpha}^2-\sigma_{\beta}^2}
    {\sigma_{\alpha}\sigma_{\beta}}. \label{f_2_aniso_diff}
\end{equation}
The latter is symmetric with respect to the reversal of $\alpha$ and
$\beta$. Finally, combining
(\ref{C_near_pc},\ref{C_slow},\ref{delta_1},\ref{f1_xi},\ref{c}), we
write the anisotropic solution:
\begin{equation}
    C_{\alpha}(q) = R + \frac{2}{z-2} \xi(\sigma_{\beta}\epsilon q)
    \epsilon
    + f_{\alpha}^{(2)}(\sigma_{\beta} \epsilon q) \epsilon^2.
\end{equation}
Then, using (\ref{b_av_def}), (\ref{sigma_avg}) and
(\ref{n_alpha_n_beta}), we obtain the average conductivity up to and
including $\epsilon^3$ order:
\begin{eqnarray}
    \overline{\sigma_{\alpha}}_{\text{(II)}} &=& \frac{z\sigma_{\alpha}}
    {\sigma_{\alpha}+\sigma_{\beta}} \left\{ 0.762  \frac{2}{z-2} \sigma_{\beta}
    \epsilon^2 - \sigma_{\beta} f_{\alpha}^{(2)'}(q) \epsilon^{3}
    \right\} \nonumber \\
    &&+ C\epsilon^{3}, \label{sigma_avg_exp_crit1}
\end{eqnarray}
where $C$ is a constant which could be determined by equating the
terms proportional to $\epsilon^3$ in the expansion of eq.
(\ref{Int_eq_near_pc_1}). As a result the first term in the
expansion (\ref{sigma_avg_exp_crit1}), proportional to
$\epsilon^{2}$, is the symmetric one. In addition, we can calculate
the difference in the derivatives of $f^{(2)}_{\alpha}$ and
$f^{(2)}_{\beta}$ using  eqs. (\ref{f_2_aniso_diff},\ref{xi}), which
yields
\begin{equation}
    \overline{\sigma_{\alpha}} - \overline{\sigma_{\beta}} =
    0.762 \frac{2(z-1)}{z-2}
    (\sigma_{\alpha}-\sigma_{\beta}) \epsilon^{3}.
\end{equation}
Therefore, we have
\begin{equation}
\frac{\overline{\sigma_{\alpha}} -
\overline{\sigma_{\beta}}}{\overline{\sigma_{\beta}}} =
    \frac{z-1}{z}\frac{\sigma_{\alpha}^{2}-\sigma_{\beta}^{2}}
    {\sigma_{\alpha}\sigma_{\beta}}\epsilon.
    \label{aniso_ration_app}
\end{equation}

\end{document}